\documentclass[letterpaper, 10 pt, conference]{ieeeconf}  
\usepackage{graphicx}
\usepackage{multirow}
\usepackage{colortbl}
\usepackage{arydshln}
\usepackage{amsmath}
\usepackage[table,xcdraw]{xcolor}
\IEEEoverridecommandlockouts                              
\title{\LARGE \bf
Smart Insole: A Gait Analysis Monitoring Platform Targeting Parkinson's Disease Patients Based on Insoles}

\author{Dimitrios G. Boucharas, Christos Androutsos, George Gkois, \\ Vassilis D. Tsakanikas, Vasileios C. Pezoulas, Dimitrios Manousos, Vasileios Skaramagkas, \\ Chariklia Chatzaki, Stathis Kontogiannis, Christos Spandonidis, Alexandros K. Pantazis, \\ Nikolaos S. Tachos, Manolis Tsiknakis, Dimitrios I. Fotiadis, \emph{Fellow, IEEE.}%
\thanks{D.G. Boucharas, C. Androutsos, N.S. Tachos, G. Gkois, V.D. Tsakanikas, and V.C. Pezoulas, are with the Unit of Medical Technology and Intelligent Information Systems, Department of Materials Science and Engineering, University of Ioannina, GR-45110, Ioannina, Greece.}%
\thanks{D. Manousos is with the Institute of Computer Science, Foundation for Research and Technology Hellas (FORTH), GR-70013, Heraklion, Crete, Greece.}%
\thanks{C. Chatzaki and M.Tsiknakis are with Biomedical Informatics and eHealth Laboratory, Department of Electrical and Computer Engineering, Hellenic Mediterranean University, Estavromenos, GR-71004, Heraklion, Crete, Greece and with the Institute of Computer Science, Foundation for Research and Technology Hellas (FORTH) and the Department of Electric and Computer Engineering, Hellenic Mediterranean University, GR-71004, Heraklion, Crete, Greece.}%
\thanks{V. Skaramagkas is with the Institute of Computer Science, Foundation for Research and Technology Hellas (FORTH) and the Department of Electric and Computer Engineering, Hellenic Mediterranean University, GR-71004, Heraklion, Crete, Greece.}%
\thanks{S. Kontogiannis is with PD Neurotechnology Ltd.,  GR-45110, Ioannina, Greece.}%
\thanks{C. Spandonidis is with Prisma Electronics SA, GR-17564, P.Faliro Athens, Greece.}%
\thanks{A.K. Pantazis is with the Microelectronics Research Group, Institute of Electronic Structure and Laser (IESL), Foundation for Research and Technology Hellas (FORTH), GR-70013, Heraklion, Crete, Greece.}%
\thanks{Dimitrios I. Fotiadis is with Biomedical Research Institute, FORTH, GR-45110, Ioannina, Greece and with the Unit of Medical Technology and Intelligent Information Systems, University of Ioannina, GR-45110, Ioannina, Greece.
({\tt\small corresponding author e-mail: fotiadis@uoi.gr})}
}

\begin{document}

\maketitle
\thispagestyle{empty}
\pagestyle{empty}

\begin{abstract}
During the preceding decades, human gait analysis has been the center of attention for the scientific community, while the association between gait analysis and overall health monitoring has been extensively reported. Technological advances further assisted in this alignment, resulting in access to inexpensive and remote healthcare services. Various assessment tools, such as software platforms and mobile applications, have been proposed by the scientific community and the market that employ sensors to monitor human gait for various purposes ranging from biomechanics to the progression of functional recovery. The framework presented herein offers a valuable digital biomarker for diagnosing and monitoring Parkinson's disease that can help clinical experts in the decision-making process leading to corrective planning or patient-specific treatment. More accurate and reliable decisions can be provided through a wide variety of integrated Artificial Intelligence algorithms and straightforward visualization techniques, including, but not limited to, heatmaps and bar plots. The framework consists of three core components: the insole pair, the mobile application, and the cloud-based platform. The insole pair deploys 16 plantar pressure sensors, an accelerometer, and a gyroscope to acquire gait data. The mobile application formulates the data for the cloud platform, which orchestrates the component interaction through the web application. Utilizing open communication protocols enables the straightforward replacement of one of the core components with a relative one (e.g., a different model of insoles), transparently from the end user, without affecting the overall architecture, resulting in a framework with the flexibility to adjust its modularity. 
\end{abstract}

\begin{keywords}
Gait analysis, plantar pressure data, Parkinson’s disease, gait patterns, computer architecture
\end{keywords}

\section{INTRODUCTION}
The interest in human gait analysis has reignited during the last decades. The technological advances yielded a wide diversity of benefits for analyzing human movement that transformed enormous laboratories equipped with several video and infrared cameras, high-performance computers, and soaring-cost treadmills, force plates or walkways into low-cost, lightweight, and flexible; yet precise devices called sensors. Such devices allow individuals with limited or no access to public healthcare services to obtain a gait evaluation. Depending on the sensor placement, distinct motifs can be derived. Foot-based sensors such as pressure and Inertial Measurement Unit (IMU) sensors have been widely adopted in gait monitoring due to walking and running instabilities being heavily linked with chronic diseases, and, by extension, with overall health. Improper gait, in-addition, results in more strain on multiple body parts, which may lead to injuries.

Various assessment tools have been proposed by the scientific community and the market that exploit data derived from foot sensors. These tools have not only been utilized for sports science, biomechanics, and monitoring the progression of functional recovery, but have also expanded to be a valuable digital biomarker for diagnosing and monitoring several neurological disorders such as Parkinson's disease (PD). Furthermore, they can be employed by the general population for preventing the progression of gait inconsistencies that might lead to severe issues, if left untreated. The results acquired from the imminent gait analysis might provide the clinical expert or the physician with the information needed for corrective planning and patient-specific treatment.

PD is a neurological disorder that affects the ability of patients to control their movement. The most common symptoms are tremors (e.g., hands, legs, and jaw), stiffness of muscles, movement slowness, and sudden loss of balance and coordination. The latter may lead to falls that are the leading cause of death related to injuries among 65-year-old adults \cite{bib1, bib2}. As the disease progresses, patients face difficulties with walking and talking. Risk factors are yet to be discovered; however, it is proved that the prevalence increases with age \cite{bib3}. Currently, there is no cure for the disease, but treatments are available to relieve the patients from the symptoms and often drastically maintain, if not improve, their quality of life \cite{bib4}. Consequently, sensor-based gait analysis assessment tools can help doctors administer patient-specific therapy and corrective planning. On the other hand, patients will save time by avoiding long queues in healthcare facilities, money due to less frequent doctor visits, and energy as the data collection procedure occurs in their daily living without requiring additional effort at the clinical sites. Considering that monitoring can end up being long-term, it comes with tremendous advantages for the individuals involved.

A variety of platforms has also been proposed to offer a monitoring system that combines contextual information extracted from the Inertial Measurement Unit (IMU) and pressure sensors with information derived from gait metrics. Architectures have been constructed and refined to yield more robust results, enabling several innovative features, such as cloud computing. Over the preceding years, the examination of the gait has been aided by cloud technology. For healthcare purposes, cloud computing plays a vital role in processing large amounts of data using various decision-supporting methods. Researchers have recognized that platforms, as clinical tools, can lead to ideal gait analysis systems to observe abnormalities and identify the induced severity. But their role is not limited to that, but further expanded to offering a complete view of the patients' gait parameters by visualizing specific pressure diagrams and IMU graphical representations, among others. The approach, as described above, goes beyond the traditional healthcare approaches defining a new entry in the healthcare era named Telehealth \cite{tele}.

Pressure sensors, accelerometers, and gyroscopes are the most widely deployed gait-based sensors, while their placement varies on different sites of the lower human body. For more reliable and accurate analysis, the sensors should be employed in the shape of shoes, insoles, or outsoles. The latter captures human motion and, by extension, movement patterns. The exploitation of cloud services further supports the analysis with efficient resource allocation, along with quick configuration and implementation. The collected raw data are transmitted, stored, and analyzed by a pool of Machine Learning (ML) or Deep Learning (DL) algorithms in the cloud. The effectiveness of this procedure can potentially create a pathway for remote health monitoring that will allow healthcare professionals such as clinicians, physicians, and formal caregivers to continuously supervise and evaluate a person's health data in real-time, even from distant locations.

\section{Related Work}
According to the literature, several gait analysis platforms have been developed, presented, and evaluated. Ziagkas et al. introduced a portable, yet low cost tool for gait analysis, in daily living by synthesizing a walking profile \cite{bib5}. The latter features several gait metrics based on data derived from a pair of smart insoles where integrated sensors are deployed. A summary of the gait analysis results is presented in a graphical manner on a platform utilizing an intermediate bluetooth connection box. To validate the underlying architecture, a human motion-capturing ecosystem composed of high-resolution infrared cameras, a central unit that acts as a camera synchronizer, and a processing unit responsible for Vicon software execution was constructed. The results indicate that some divergence can be observed between the two systems, mainly affected by the low number of participants and the extracted gait parameters; however, the provided measurements are valid. Loukovitis et al. extended the previously described work by exploring the repeatability of the developed system \cite{bib6}. The findings, extracted from two distinct trials conducted by 22 participants, support the reliability of the system in analyzing gait parameters based on the intraclass correlation coefficient.

Chen et al. proposed a gait monitoring framework based on piezoelectric insoles focused on three distinct target groups suffering either from PD, stroke or diabetes \cite{bib7}. These chronic diseases, along with their progression, could possibly result in functional gait inconsistencies. 
The developed framework employs an Internet of Things approach to connect and exchange gait data between heterogeneous devices and systems over the Internet. In essence, the smart insole connects to the user's computer via bluetooth, and the Center of Pressure (COP) data are uploaded to a cloud server from where a clinical institution can retrieve it for medical purposes. 
Consequently, the presented framework offers great convenience to both patients and doctors by avoiding frequent visits to the hospital, leading to a complete solution.

Tsiouris et al. synthesized a mobile healthcare platform that acts as a complete assistant targeting PD patients \cite{bib8}. The accelerometer, gyroscope, microphone, and pressure sensors were integrated into the proposed solution to capture a wide variety of gait parameters to evaluate the severity of symptoms caused by the disease. The sensors were placed on a commercial wrist, an insole pair, and a mobile device. The latter was also used to estimate the patient's cognition, mental state, and diet through a series of tests that resulted in a review of motivations and obliging recommendations. Patient data was uploaded to the cloud encrypted, where Artificial Intelligence (AI) methods were recruited to estimate the symptoms and progression of the disease. Another mobile application was designed and developed for direct access to patient data and real-time monitoring by clinicians. Clinicians might assess the information derived from the cloud-embedded algorithms to adjust the duration and dosages of the treatment accordingly. The system also included a mechanism to not only notify clinicians when new symptoms appeared or existing ones worsened, but also to suggest which course of medication modifications to take.

PD is difficult to diagnose because, at present,

\begin{enumerate}
  \item it is based on a patient's clinical evaluation \cite{bib8.5};
  \item although there are some promising leads, there are no definitive biomarkers for PD, and all relevant research findings work supportively for the confirmation of the clinical diagnosis;
  \item the symptoms develop gradually, with a lengthy delay between the actual degeneration of dopaminergic neurons (70-80\% loss) and the start of recognizable clinical symptoms \cite{bib8.6};
  \item the variety of symptoms and the heterogeneity of their onset and progression contribute to the complexity of PD clinical phenotypes \cite{bib8.7};
  \item an extra challenge to the diagnosis is accumulated due to the fact that several syndromes mimic the symptoms of PD.
\end{enumerate}

As a consequence, the evaluation of the disease severity, which plays a vital role in the progression, is also hard to be addressed, especially in the early stages fo the disease. To estimate the stages of PD, Balaji et al. compared several ML models that could help make an accurate diagnosis. Based on the Unified PD Rating Scale (UPDRS) and the Hohen and Yahr Scale (HY), 
a classification problem was devised \cite{bib9}. The dataset was based on vertical ground reaction forces (VGRF) derived from PD patients scoring above 2.0 in HY scale. 

\begin{figure*}[!hbt]
	\centering
	\includegraphics[width=17cm]{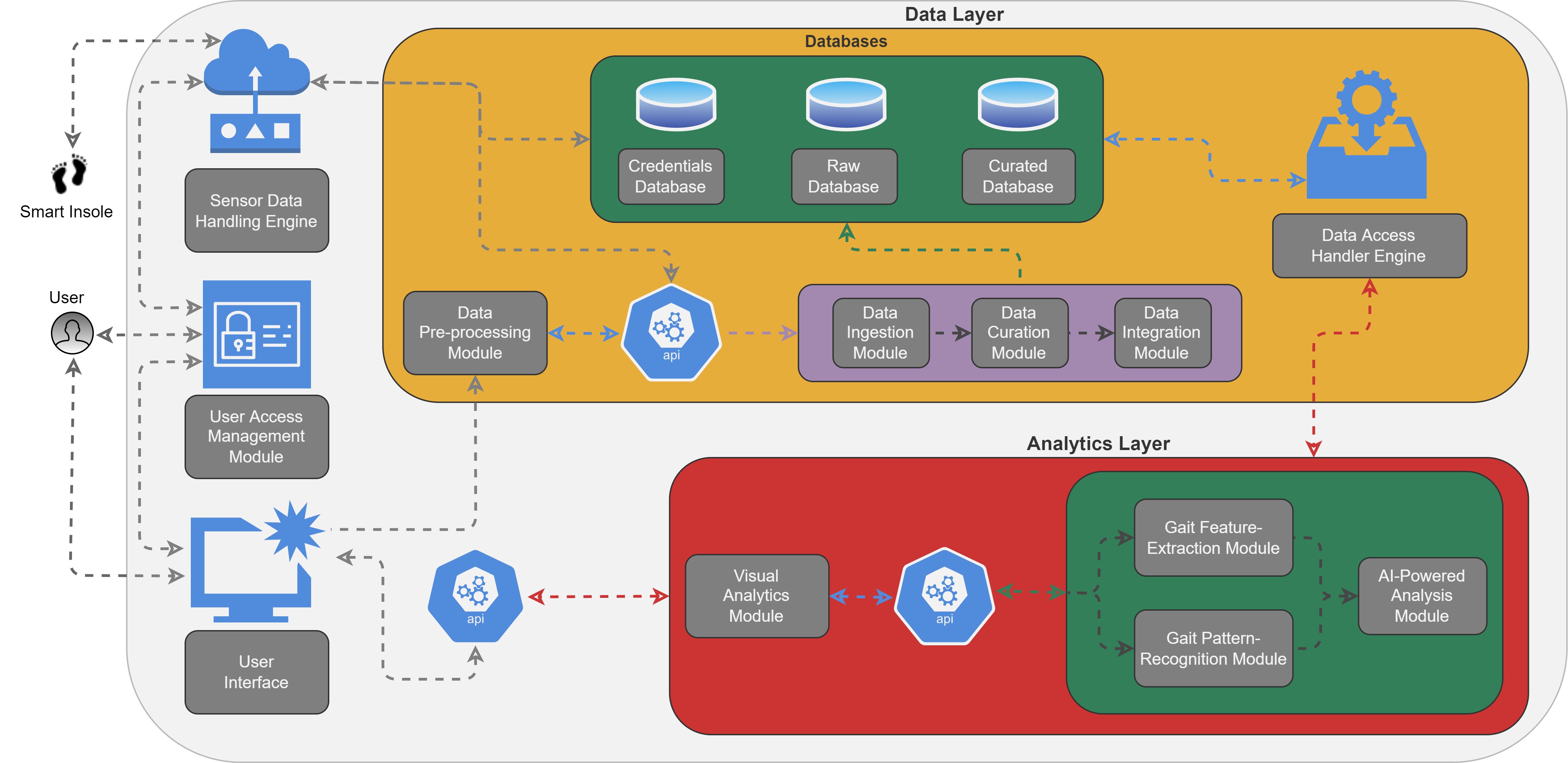}
	\caption{The 'Smart Insole' Architecture}
	\label{fig1}
\end{figure*}

The feature vectors were constructed with gait spatio-temporal parameters (e.g., cadence, step length, and swing time) and fed into four distinct classifiers. Feature selection techniques were employed to improve the performance and robustness of the developed models. The classifiers employed were Decision Trees (DT), Support Vector Machines (SVM), an Ensemble Classifier (EC), and a Bayes Classifier (BC), with DT producing the most promising results.

Lazzaro di Biase et al. attempted to summarize the discriminative gait features that might assist in distinguishing PD patients from healthy subjects, along with the PD stages \cite{bib10}. In early stages, gait observations exhibit increased disparity reflecting instability, a few spatio-temporal parameters demonstrate some degree of deduction, such as the reduced step length, and dual-task sessions (e.g., walking and counting backwards) are hard to comprehend for the PD patients. In mild to moderate stages, body asymmetry diminishes. Accordingly, the double support time increases. In more advanced stages of PD, the gait further worsens with increased frequency in freezing of gait (FOG), loss of balance and lack of coordination.

Targeting the ageing population, Garc{\'\i}a-Villamil et al. proposed a wearable gait assessment device based on a 6-axis inertial sensor \cite{bib10a}. The sensor data, collected across different surfaces such as indoor and outdoor flooring, were utilized to measure several gait parameters. The system includes a mobile application that visualizes gait-related information for users. Device reliability was validated employing the intraclass correlation coefficient, which was found to be 0.69. The relationship between the gait characteristics obtained by the device and clinical tests was also explored. Sequentially, a close connection was detected between some gait parameters (e.g., mean speed, cadence, and mean stride) and the short physical performance battery test, which combines gait speed, chair stand, and balance tests to produce a frailty score. The results showed that frailty and falls could be predicted accurately upon utilizing the score in the ageing population. To the same extent, Aspera et al. examined the ability of gait parameters to approximate the levels of frailty (for example, frail, prefrail, robust) presented by the elderly population \cite{bib10b}. Wireless inertial sensors were placed on various parts of the lower human body to record the corresponding gait data streams. Logistic models were recruited to explore associations between frailty and the gait parameters extracted. Receiver operating characteristic curves were employed to calculate the area under the curve for each parameter, while the Youden index was considered for the cut-off values. The results indicate that the gait parameters could distinguish frailty levels, especially between frail and robust levels.

\section{Our Contribution}
To address the aforementioned needs, an advanced wearable solution has been developed. The 'Smart Insole' \cite{bib-1} is an innovative digital ecosystem, consisting of a pair of smart insoles (integrating various sensors), a mobile application and a cloud-based platform. Specifically, 16 distinct plantar pressure sensors, an accelerometer, a gyroscope, and a magnetometer were embedded to the insole for acquiring gait raw data. A mobile application was designed and developed to receive the insole data wirelessly (via BLE) and provide valuable gait-related information to the end user (PD patients and elders). 
The 'Smart Insole' cloud platform encompasses the web services, the well-defined components, and their interaction. 
The developed and integrated web application is responsible for data ingestion, secure data storage, data curation, and data integration, among other modules. The 'Smart Insole' platform also incorporates an AI-based reasoning engine to provide decision support to the clinical experts. The work presented herein describes in detail the architecture and functionalities provided by the 'Smart Insole' solution. 
The proposed solution serves as a supplemental expert support to align into a more reliable, accurate, and patient-specific decision-making process based on graphical representations, understandable motifs, and gait characteristics that are not limited to features derived from gait cycle phases from older individuals and patients with PD. 

\section{ARCHITECTURE}

The proposed architecture, depicted in Fig. \ref{fig1}, contains three core components: the smart insoles, the cloud-based platform, and the mobile application. The system targets elderly population and patients with PD who can use smart insoles in their daily living, together with the corresponding mobile application to obtain informative gait-related reports. It is worth mentioning that the proposed architecture was designed in order the system to be independent of the Smart Insole characteristics. The system can support third-party insoles on top of the developed ones, without compromising the performance of the whole system. Specifically, the sensor data handling engine is responsible for mapping the incoming raw data to the Smart Insole data model and performing any normalization to meet the specifications for the data analysis layer. 
The cloud-based application has as end users the healthcare professionals by providing to them rich and informative clinical reports for the gait analysis. These reports provide gait-related data, metadata visualizations, and comprehensive visual analytics, interlinked with gait patterns.

The architecture design encompasses all the vital elements that compose the web services and their interaction. The architecture can be hierarchically decomposed into user management-related modules, the sensor data handling engine, the data layer, and the analytics layer.

The user management-related module includes the security components (i.e., the SSL/TLS protocols and the OAUTH2 framework) that are invoked during user authentication and user access management procedures and use any necessary resources from the hardware infrastructure and software components, as well as from the private database of user credentials for user authentication.

The engine is designed as a three-node pipeline including the data ingestion, data curation, and data integration module. The sensor data handling engine is responsible for collecting the sensor data from the insoles and directing the data to the REST API where the pre-processing (e.g., curation) takes place.

Next, the data layer includes dedicated databases where the data storage occurs, from where the data access handler engine will enable the AI-powered analysis module. Data storage is characterized by various data storage technologies, including NoSQL, SQL, and a file storage system. AI-powered analysis module utilizes advanced AI models and algorithmic pipelines to assess data information extracting valuable decision support tools, such as graphs and metrics. The visual analytics module will then attempt to visualize these graphs to the user interface through the rest API, once the end user logins successfully by providing the credentials. REST APIs provide an interface to all available system functions and data. All data, either received or requested through the REST API, are passed to the data scheduler, which is responsible for forwarding to the corresponding functional components of the system.

\subsection{THE SMART INSOLES}

\begin{figure}[!hbt]
	\centering
	\includegraphics[width=8cm]{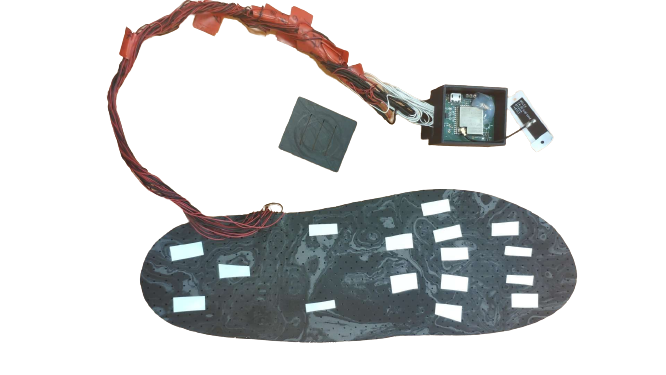}
	\caption{The Smart Insoles}
	\label{fig2}
\end{figure}

The insoles, as depicted in Fig. \ref{fig2}, are one of the main components of the system which provide the plantar pressures and inertial unit (IMU) data to monitor the gait spatio-temporal characteristics. The pair of insoles integrates a number of piezoresistive sensors to assess the pressure exerted by different parts of the foot plantar during the gait cycle. The IMU sensor includes an accelerometer, gyroscope, and magnetometer to capture the velocity and orientation of the end-user foot during the various gait phases. The insoles include an electronic component that acts as a communication gateway. This subsystem incorporates a microprocessor that collects data from the sensors and assigns them acquisition timestamps. The system connects wirelessly with a mobile device via a Bluetooth Low Energy (BLE) protocol to transfer the gait-related data. The Smart Insole prototype integrates 16 pressure sensors which are manufactured exploiting 3D printing technology. 



\subsection{THE MOBILE APPLICATION}

In the 'Smart Insole' system, a mobile device acts as a gateway to collect data from smart insole devices via BLE and send it to the cloud server for storage and analysis. A preprocessing step lies between these steps to formulate the data into an appropriate structure that the cloud platform demands. The mobile application provides the ability to store gait-related data from smart insole devices in a local SQLite database. Synchronizing these data with the data stored on the remote server occurs when the wearable device is connected to the Internet (via WiFi or the 4G/5G network). The application is responsible for pairing with the smart insole device, the main component of the system, and registering a new user by creating a user account and configuring the insole settings. 

Furthermore, the mobile application provides a diverse set of capabilities for its users, such as graphical representations of postprocessed measurements, gait characteristics, and pressure distribution per activity period. 
The processed gait related data and metadata are derived from the Smart Insole cloud-based platform. Data are transferred to the cloud platform in an encrypted way safeguarded by the application.

\begin{figure}[!hbt]
	\centering
	\includegraphics[width=8.5cm]{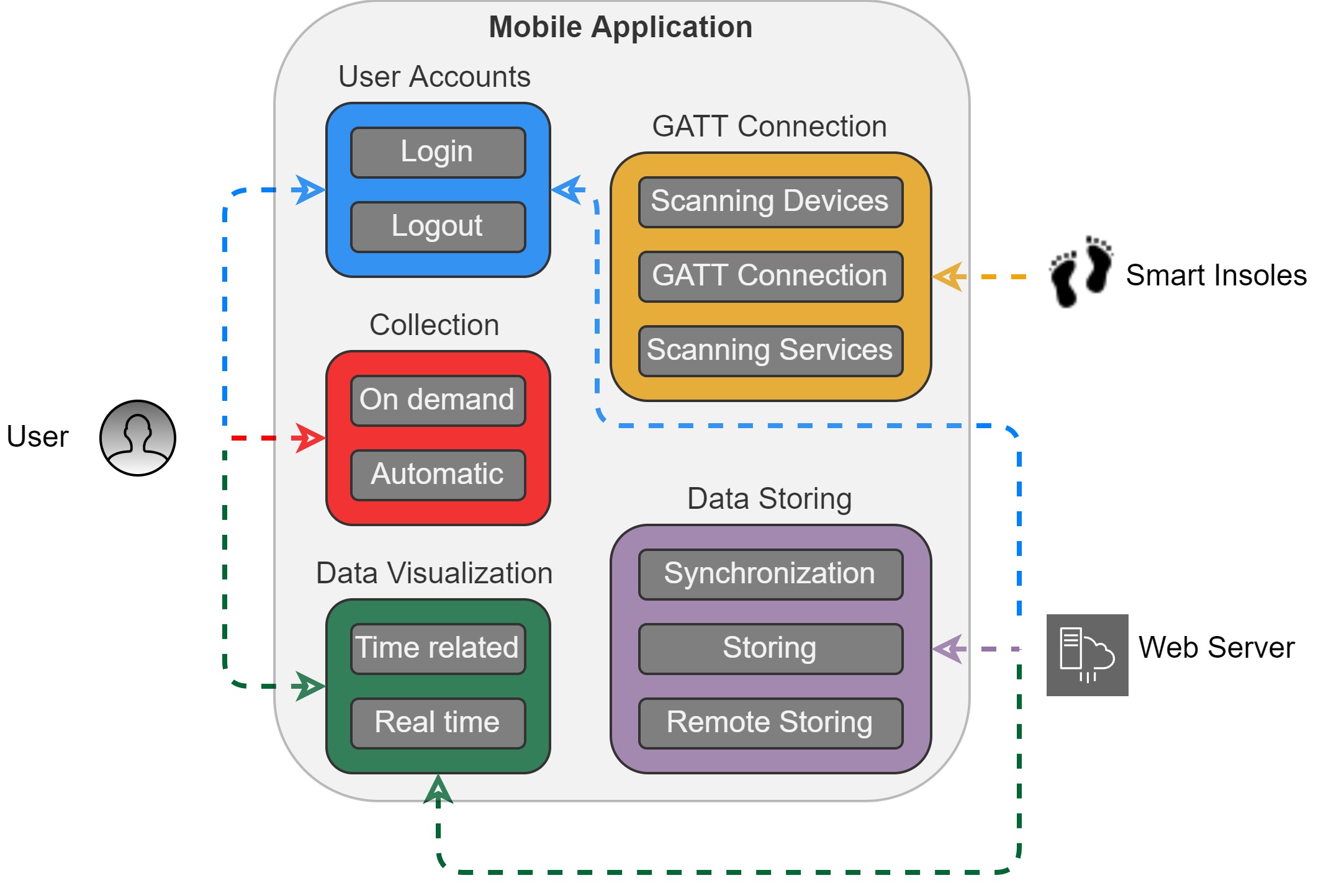}
	\caption{The Functionalities of the 'Smart Insole' mobile application}
	\label{fig3}
\end{figure}

Figure \ref{fig3} encapsulates the interactions between user, device, and web server through the mobile application. The functionalities that reside in the mobile application are user log-in and log-out, data visualization, data retrieval from insole devices, search and connect with the insole devices, remote storage (on server), and remote database synchronization.

The user must have a personal account on the application to be able to use the application. 
Upon successful login to the application, a session (user session) is created that lasts for a specified period of time, during which the user does not need to go through the login process again. 

\begin{figure}[!hbt]
	\centering
	\includegraphics[width=8.5cm]{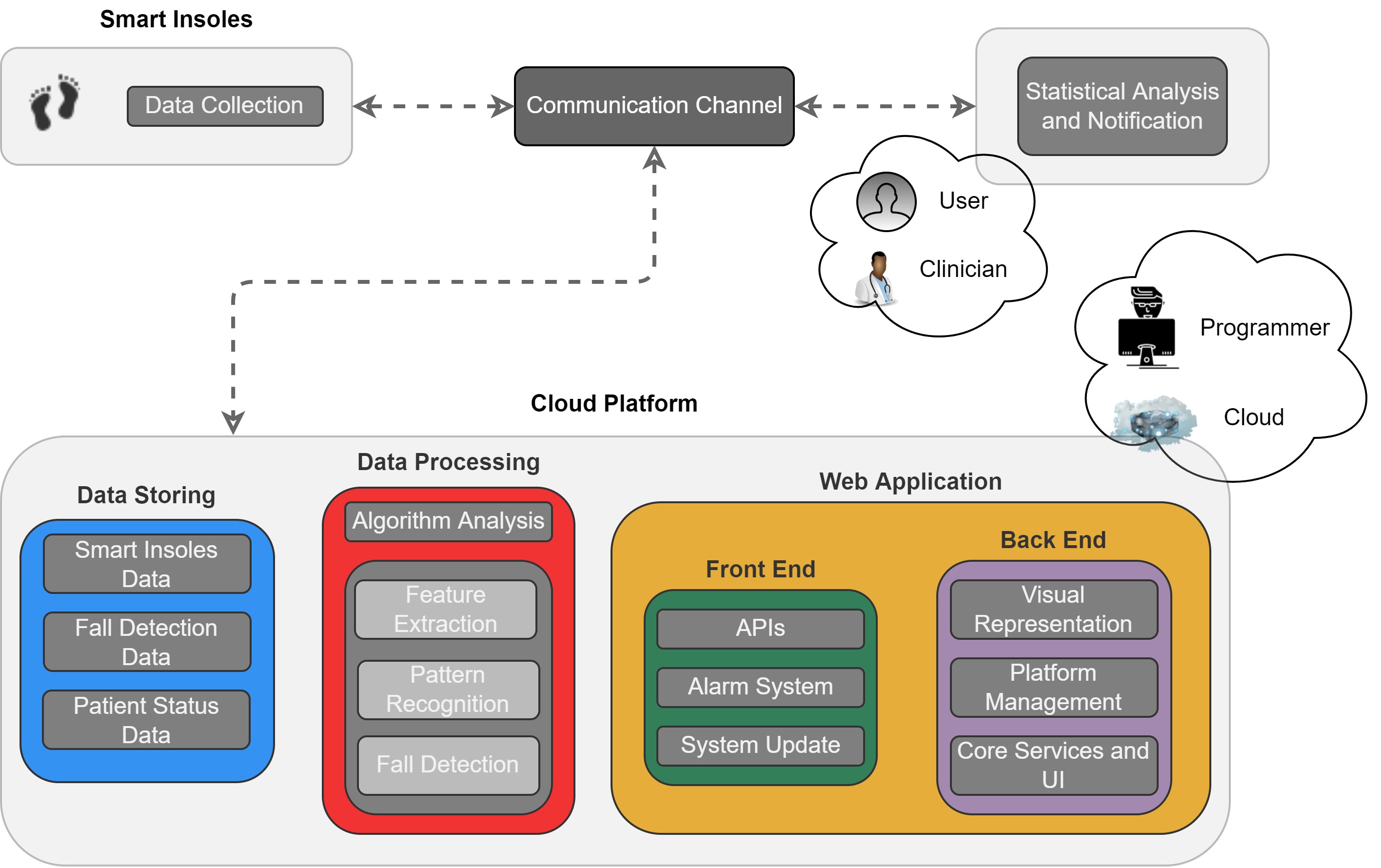}
	\caption{The Interactions of the Smart Insole Components}
	\label{fig4}
\end{figure}

Once the data collection procedure initiates, a process is performed to automatically update the application with each new measurement. Every measurement taken by the insole devices, is stored by the application in the local database. The synchronization process with the cloud infrastructure depends on the existence of an Internet connection. 
The local database is used to temporarily store raw data retrieved from the insole devices until the mobile application can connect to the cloud-based platform, where the data are permanently stored. 
To visualize the gait data, the GraphView library was utilized through different types of graphs such as line graphs, bar charts, scatter plots, and real-time graphs featuring scroll, scale, or zoom functionality \cite{biblib}.


\subsection{THE WEB APPLICATION}

The software of the proposed cloud technology platform is the Web application. It includes the operating engine (backend) and the services of the working environment (frontend). The back-end is responsible for all the processes executed in the background (i.e. listening and responding to requests), while the front-end is dedicated to user interaction such as user management and patient monitoring. It is also responsible for API requests, data storing and retrieving, and user authentication providing an easy and quick expansion to the system. 

\begin{figure}[!hbt]
	\centering
	\includegraphics[width=8.5cm]{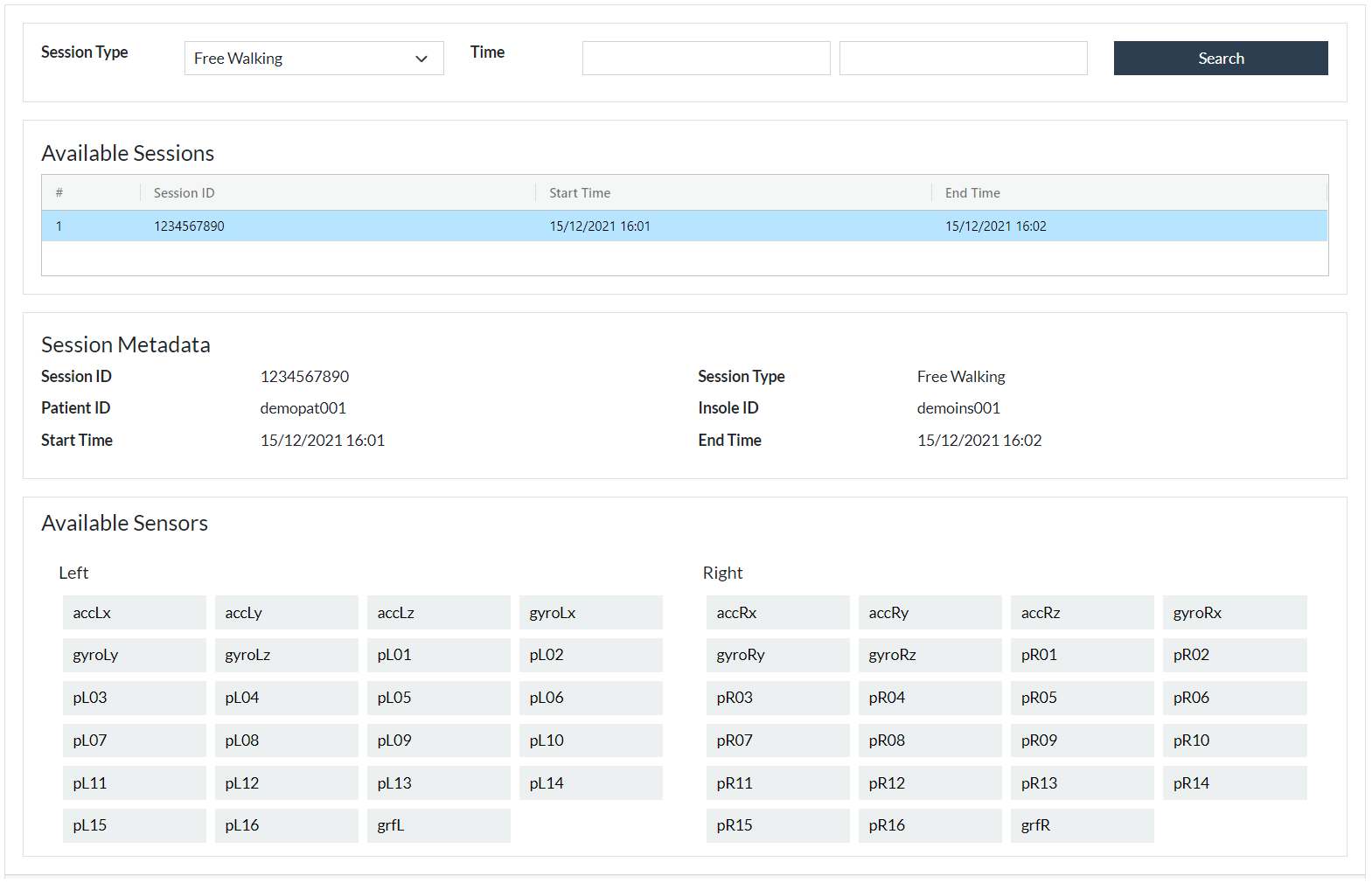}
	\caption{The insole raw sensor data report page.}
	\label{fig5}
\end{figure}

Specifically, several databases can be connected to this service, while it can serve many and varied types of application. The primary benefit of the application is the access control, which can be easily adapted to any application connected to the service. Utilizing the service can reduce the application workload caused by requests compared to applications employing their own database connection and management schemes.

\begin{figure}[!hbt]
	\centering
	\includegraphics[width=8.5cm]{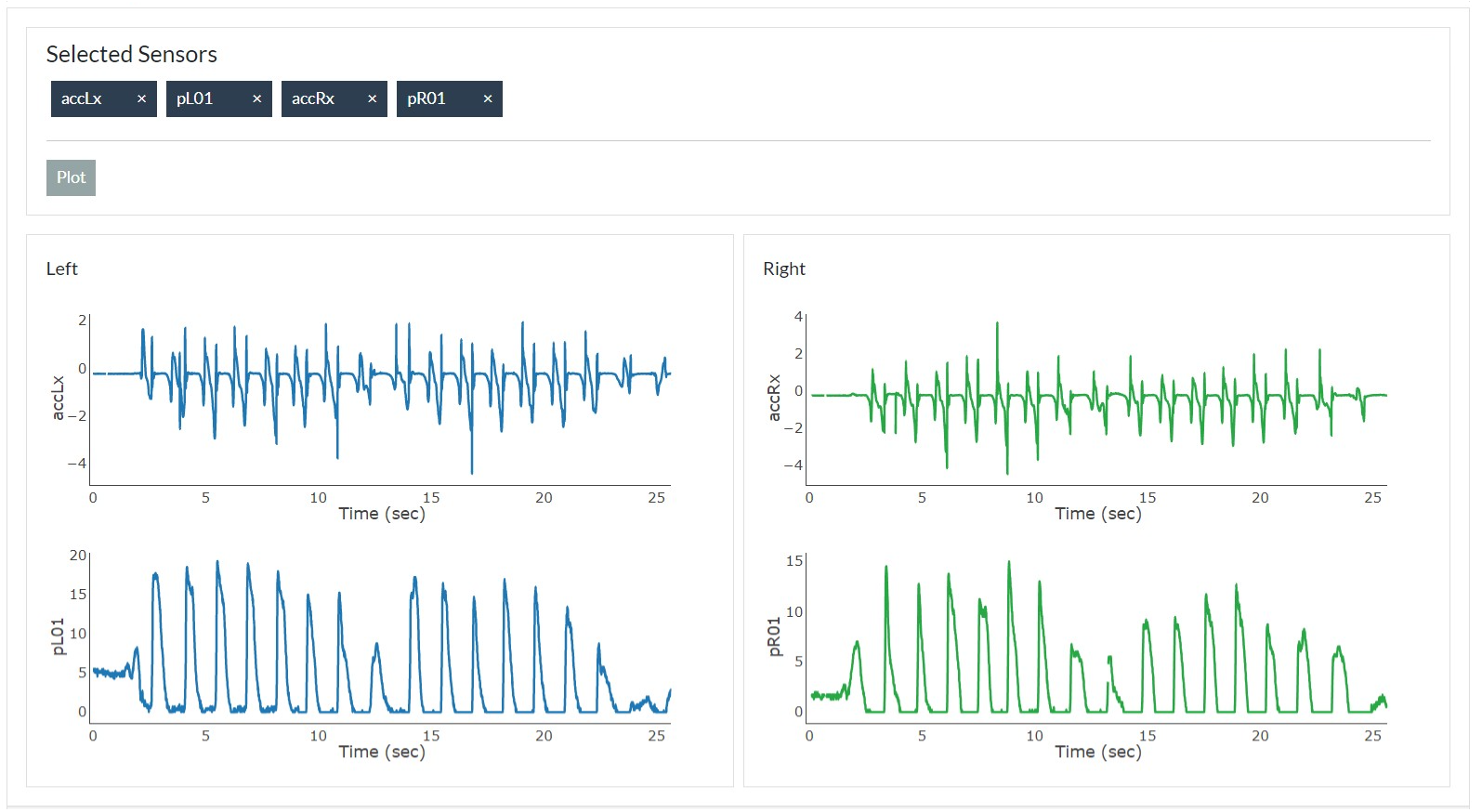}
	\caption{The insole raw data visualization.}
	\label{fig6}
\end{figure}

In essence, the web application provides several analytical summary reports based on the type of session (e.g., walking and standing balance) to the end user. Sequentially, based on the raw data acquired from free-walking, 10-meter straight walking, timed up and go (TUG) and standing tests, corresponding analytical summary reports are produced. The insole raw sensor report enables the user to choose from the patient/insole pairs by scrolling down the available list.

\begin{figure}[!hbt]
	\centering
	\includegraphics[width=8.5cm]{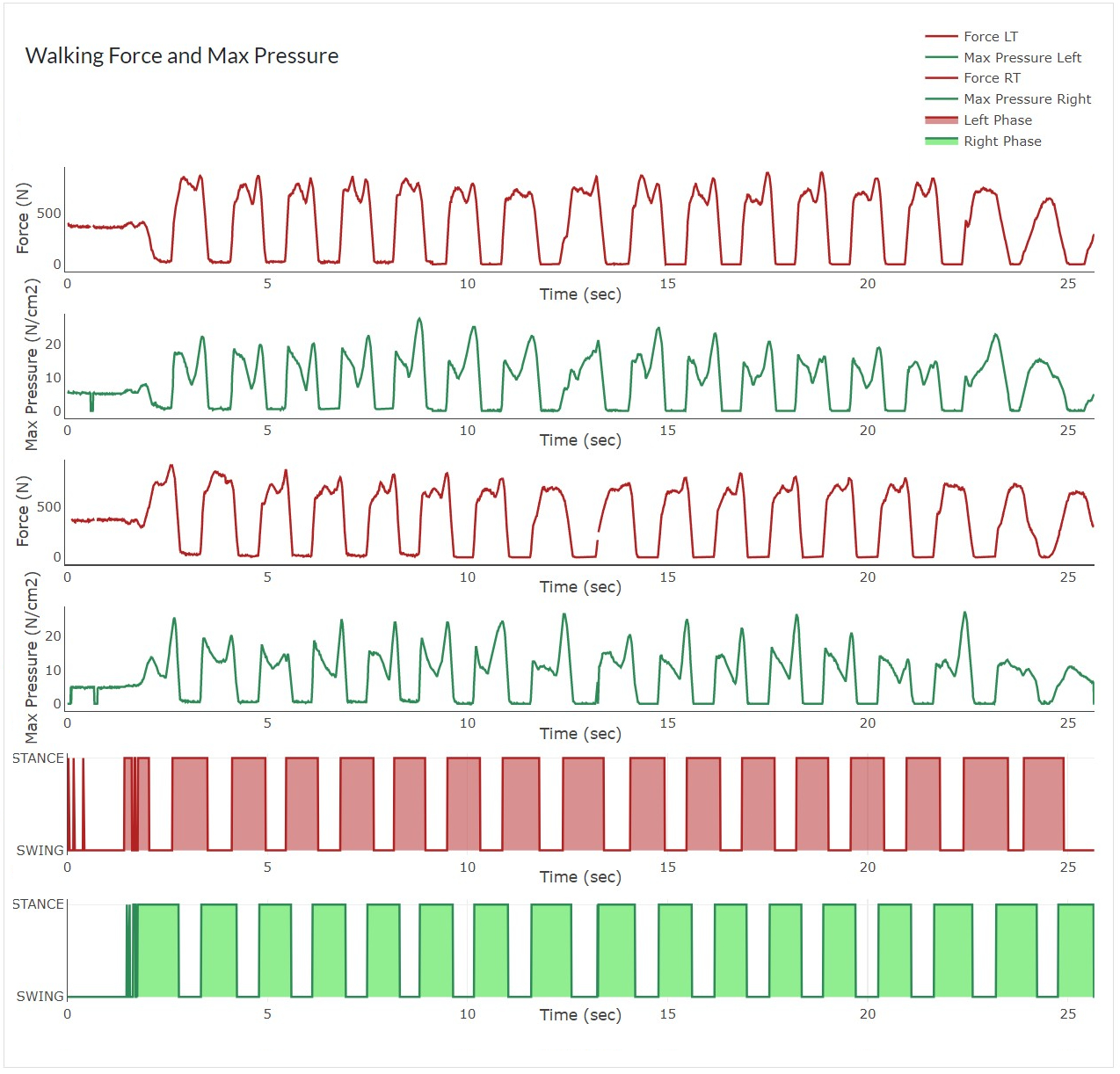}
	\caption{Visualizing force and pressure on raw sensor data.}
	\label{fig7}
\end{figure}

After selecting a pairing, the user can search for previous raw data using the integrated calendar within the time field or leave it empty to retrieve the complete history without time-related constraints. Selecting a session allows the user to choose from the pool of deployed sensors (Fig. \ref {fig5}) on the right and left foot to visualize the corresponding data. Data visualization plays a crucial and integral role in the clinician's decision by offering a way to observe the variability or divergence from the general population.

\begin{figure}[!hbt]
	\centering
	\includegraphics[width=8.5cm]{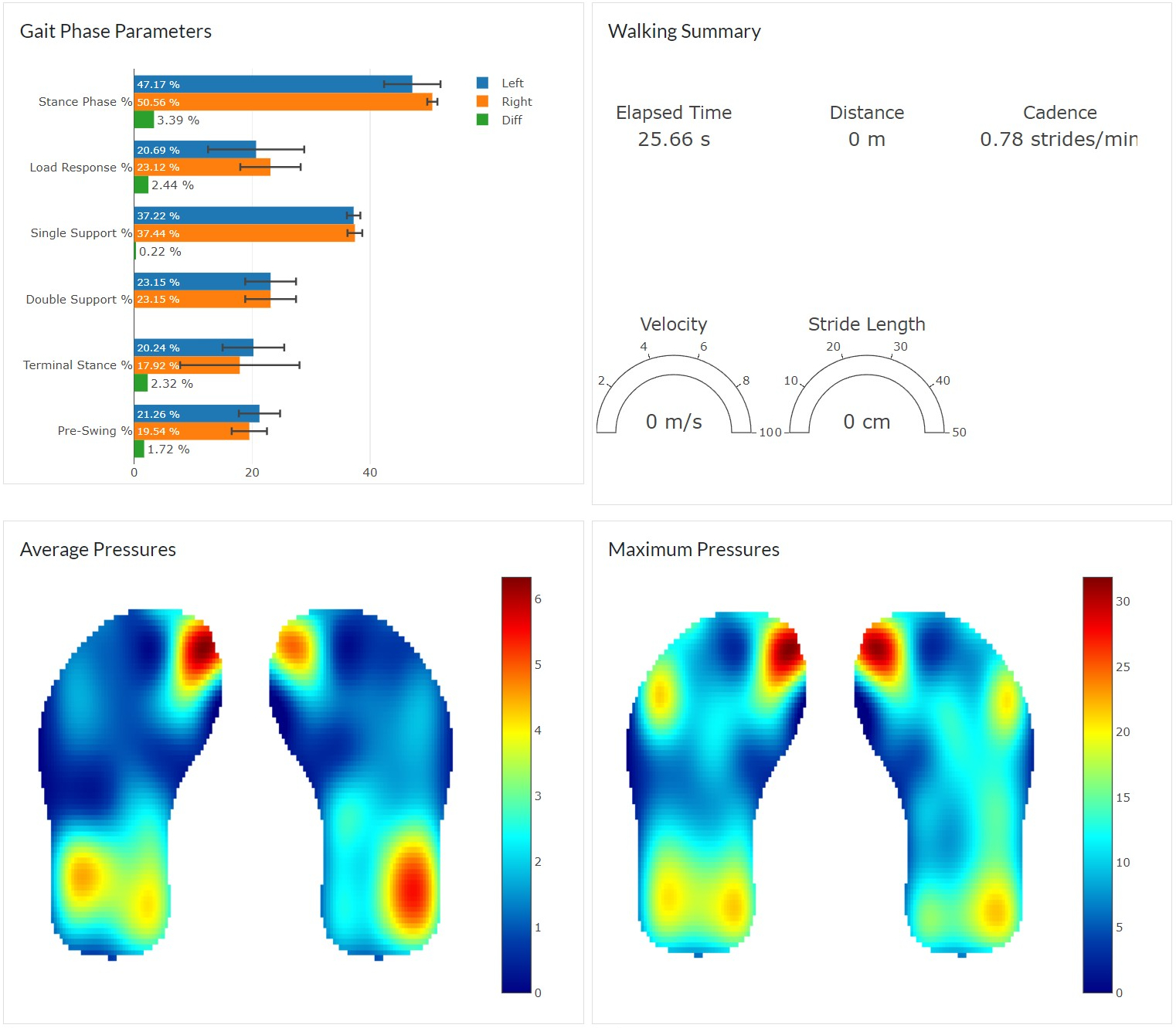}
	\caption{Gait parameter graphical representation and pressure heatmaps.}
	\label{fig8}
\end{figure}

A figure precedes (Fig. \ref{fig6}) that illustrates the progression of raw sensor data over time under the previously selected sensors, where clinical experts can inspect and provide feedback to end users. On the other hand, for free walking, straight 10-meter walking or TUG exercises, the system offers automated detailed walking summary reports consisting of several gait parameters: single support time, double support time, cadence, stance phase, pre-swing, cycle time, load response, and terminal stance \cite{tug}.

The mean value, standard deviation, and respective p values are also calculated per gait parameter to synthesize box plots, while line plots are utilized for graphical representations of force and stance (Fig. \ref{fig7}). Furthermore, pressure heatmaps and center of pressure plots are recruited to assess the plantar pressure distribution as depicted in Fig. \ref{fig8}. In conclusion, standing balance reports consist of sway analysis graphs representing medial-lateral and anterior-posterior deviation per foot, center of pressure plots, and butterfly graphs. The motifs drawn by the conducted analysis and the accompanying diagrams are thoroughly detailed in a prior study \cite{eai22}. 

\subsubsection{SMART INSOLE DECISION SUPPORT TOOLS}

The decision support tools consist of two categories. The first category includes a powerful computational and visualization engine, which extracts spatiotemporal metrics and renders visual representations like scatter, box and violin plots. These algorithmic pipelines and graphical representations are able to interpret even the tiniest abnormalities resulting from improper gait behavior. The second category consists of a set of clinical reports generated by AI. Due to its modularity, the specific tool can easily host other AI models, even though the currently deployed ones pertain to PD.

\begin{figure}[!hbt]
	\centering
	\includegraphics[width=8.5cm]{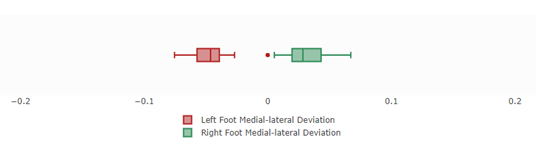}
	\caption{Sway analysis based on Medial-lateral deviation.}
	\label{fig9}
\end{figure}

The 'Smart Insole' decision support tools exploited the 'Smart Insole' dataset (Protocol approval \textit{279/14-05-2021} from Ethical Committee of General University Hospital of Patras) to develop the AI models \cite{chatza}. The dataset, which contains contextual information on acceleration, orientation and plantar pressures, was collected while healthy adults, the elderly, and PD patients completed three to five different exercise protocols. In the first exercise, participants are asked to walk 10 meters with a 180-degree turn at three speeds (e.g., slow, normal, high), whereas in the second exercise, also known as the timed up and go test, they begin seated and repeat the first exercise. Participants are instructed to walk at a normal pace. In the third test, subjects' balance is evaluated while standing for 10 seconds with their eyes open, followed by 10 seconds with their eyes closed, with their feet approximately 30 cm apart.

With regard to the presented version of the system, there are two AI models developed and integrated. The first one is a binary classifier, build on the XGBoost algorithm, which classifies an individual as a PD patient of a non-PD patient. The algorithm performance exhibits good results, yielding 0.75, 0.79, 0.65, and 0.71 in accuracy, precision, recall, and f1 score, respectively. This model can be used for screening, raising flags for individuals with "abnormal" gait profile, and suggesting a full scale assessment by the expert.


The second AI model refers to the assessment of the UPDRS scale part 3.10 related to gait. The model utilizes as input the gait features extracted from the raw data of the 10-meter walking predefined exercise and classifies the patient in a severity scale of 0, 1, 2 and 3. The model is build on a Random Forest classifier, with 0.86, 0.87, 0.73, and 0.81 in accuracy, precision, recall, and f1 score, respectively. The specific AI model can be utilized by the clinical experts to automatically assess PD patients regularly and provide information about the degradation of their condition, or even the assessment of pharmacological treatment and its influence on the specific scale.


As already described, the Smart Insole system is collecting spatiotemporal data either from real-world free walking or three predefined exercises, which are well known in clinical practise. Thus, exploiting the data derived from the stationary balance sessions, a balance assessment report was designed which provides rich information along with visual patterns about the balance status of the patients. Specifically, a pipeline was developed to create specialized heatmaps based on dynamic COP calculation. During the balance session, most of the COP coordinates tend to gather around a small area, resulting in an intense heatmap region. In the same time, observations may exhibit variability, which leads to complementary heatmap regions described as distinct points. 



The gait related reports of the system also include butterfly diagrams which were utilized to monitor the gait cycle by capturing the COP shifting patterns during walking. The annotated foot states (e.g., heel strike, heel rise, toe off, and foot flat) derived from the developed state machine contributed to the butterfly diagram construction. The height, the passing straight lines, and the symmetry of the trajectory are unique traits found in the diagrams that are explored for their contribution to differentiating PD patients from healthy individuals.


Some gait features may interpret walking session instabilities to a greater extent than others. Stance, swing, single, and double support phases can capture gait-related time distributions. For instance, the single support phase refers to the normalized time span a foot remains on the ground, while the time both feet stand on the ground attributes to the double support phase. Undoubtedly, an individual with a relevant high proportion in the double support phase faces issues with walking, possibly due to fear of falling or the need to have a higher level of gait control. Hence, the distributions are indirect indicators of a faulty or normal gait cycle. As a result, in the Smart Insole web application and specifically for the gait-related session report, circular diagrams were developed to visualize the gait-phases patterns. The latter may contribute to the decision-making from the clinical expert in the gait analysis process. 



It is evident from the data gathered over the course of the Smart Insole project that specific motifs can be drawn from sensors embedded into a pair of insoles that are capable of capturing both gait and balance patterns. If the work is developed further, advanced patterns with the potential to explain a method for detecting PD may emerge.


\section{DISCUSSION \& CONCLUSION}

The framework presented herein acts as a decision support tool for clinical experts who assess the gait quality in the population suffering from gait inconsistencies, such as the elderly and PD patients. It is not only the powerful computational and visualization engine that extracts features and metrics and visually renders them to assist in the formation of the clinical expert evaluation, but also the advanced AI algorithms that infer the patients' gait status.

The proposed architecture consists of four core components: the smart insoles, the cloud platform, the web application, and the mobile application. Although previous studies established the stepping stones for assembling efficient gait monitoring systems, the current work technically expanded it by proposing an architecture design characterized by modularity. Contrary to traditional schemes where the data acquisition component is the main parameter of the system, the Smart Insole framework is decentralized and offers a flexible solution. Hence, the developed architecture and the employed communication protocols allow the replacement of the insoles (e.g., a core component of the system) without compromising its overall efficiency, functionality, and architecture. The proposed design also takes advantage of the power of the cloud for storing, processing, and authenticating purposes, among others, resulting in advanced resource management.

Regarding the future of health monitoring platforms, multiple modalities can be exploited to synthesize an advanced Telehealth monitoring framework which is not limited to considering gait-related metrics. A framework that takes into account emotional, physical activity, mental empowerment, psychological stability, and nutritional well-being may result in complete monitoring of human health and solving equally important problems.

\section{Acknowledgement}
This research was funded by the European Regional Development Fund of the European Union and Greek national funds through the Operational Program Competitiveness, Entrepreneurship, and Innovation, under the call RESEARCH–CREATE–INNOVATE (project code: T1E$\Delta$K-01888). The article reflects only the authors’ views. The European and Greek Commissions are not responsible for any use that may be made for the information it contains.


\end{document}